\documentstyle[aps]{revtex}

\begin{document}
\def\gsim{\hbox{$\lower1pt\hbox{$>$}\above-1pt\raise1pt\hbox{$\sim$}$}}
\def\lsim{\hbox{$\lower1pt\hbox{$<$}\above-1pt\raise1pt\hbox{$\sim$}$}}
\title{
Spin susceptibility and magnetic short--range order in the
Hubbard model}
\author{{\sc U. Trapper$^{(1,2)}$}, {\sc  D. Ihle$^{(1)}$}
and {\sc H.~Fehske$^{(2)}$} \\[0.5cm]
$^{(1)}$Institut f\"ur Theoretische Physik, Universit\"at Leipzig,
D--04109 Leipzig, Germany\\[0.3cm]
$^{(2)}$Physikalisches Institut, Universit\"at Bayreuth,
D--95440 Bayreuth, Germany\\[0.6cm]}
\date{Bayreuth, April 18, 1996}
\maketitle
\newcounter{saveeqn}
\newcommand{\alpheqn}{\setcounter{saveeqn}{\value{equation}}%
\addtocounter{saveeqn}{1}%
\setcounter{equation}{0}%
\renewcommand{\theequation}{\arabic{saveeqn}.\alph{equation}}}
\newcommand{\reseteqn}{\setcounter{equation}{\value{saveeqn}}%
\renewcommand{\theequation}{\arabic{equation}}}
\newcommand{\gapro}
  {\raisebox{-0.25ex} {$\,\stackrel{\scriptscriptstyle>}%
    {\scriptscriptstyle\sim}\,$}}
\newcommand{\lapro}
   {\raisebox{-0.25ex} {$\,\stackrel{\scriptscriptstyle<}%
    {\scriptscriptstyle\sim}\,$}}

\def\cZ{{\cal Z}}
\def\cD{{\cal D}}
\def\cL{{\cal L}}
\def\cS{{\cal S}}
\def\cH{{\cal H}}
\def\cG{{\cal G}}
\def\cF{{\cal F}}
\def\cJ{{\cal J}}
\def\cE{{\cal E}}
\def\cP{{\cal P}}

\def\ecis{c_{i\sigma}^{\dagger}}
\def\ecjs{c_{j\sigma}^{\dagger}}
\def\cis{c_{i\sigma}^{ }}
\def\cjs{c_{j\sigma}^{ }}
\def\ccis{c_{i\sigma}^{*}}
\def\ccjs{c_{j\sigma}^{*}}

\def\tecis{\tilde{c}_{i\sigma}^{\dagger}}
\def\tecjs{\tilde{c}_{j\sigma}^{\dagger}}
\def\tcis{\tilde{c}_{i\sigma}^{ }}
\def\tcjs{\tilde{c}_{j\sigma}^{ }}
\def\tccis{\tilde{c}_{i\sigma}^{*}}
\def\tccjs{\tilde{c}_{j\sigma}^{*}}

\def\niu{n_{i\uparrow}^{ }}
\def\nid{n_{i\downarrow}^{ }}
\def\nis{n_{i\sigma}^{ }}

\def\eei{e_i^{\dagger}}
\def\cei{e_i^{*}}
\def\ei{e_i^{ }}
\def\edi{d_i^{\dagger}}
\def\cdi{d_i^{*}}
\def\di{d_i^{ }}
\def\epis{p_{i\sigma}^{\dagger}}
\def\cpis{p_{i\sigma}^{*}}
\def\pis{p_{i\sigma}^{ }}
\def\epins{p_{i{-\sigma}}^{\dagger}}
\def\cpins{p_{i{-\sigma}}^{*}}
\def\pins{p_{i{-\sigma}}^{ }}
\def\ezis{z_{i\sigma}^{\dagger}}
\def\czis{z_{i\sigma}^{*}}
\def\zis{z_{i\sigma}^{ }}
\def\ezjs{z_{j\sigma}^{\dagger}}
\def\czjs{z_{j\sigma}^{*}}
\def\zjs{z_{j\sigma}^{ }}
\def\cdd{d_i^* d_i^{ }}

\def\mb{\bar{m}}
\def\xib{\bar{\xi}}

\def\si{{s_i^{ }}}
\def\sj{{s_j^{ }}}

\def\alpi{{\alpha_i^{ }}}
\def\alpj{{\alpha_j^{ }}}

\def\mbsi{\bar{m}_{\si}^{ }}
\def\xibsi{\bar{\xi}_{\si}^{ }}
\def\dsi{d_{\si}^{ }}
\def\cdsi{d_{\si}^{*}}
\def\nsi{n_{\si}^{ }}
\def\nusi{\nu_{\si}^{ }}
\def\qsi{q_{\si\sigma}^{ }}

\def\mba{\bar{m}_{\alpha}^{ }}
\def\xiba{\bar{\xi}_{\alpha}^{ }}
\def\xibab{\bar{\xi}_{{-\alpha}}^{ }}
\def\da{d_{\alpha}^{ }}
\def\cda{d_{\alpha}^{*}}
\def\cdada{d_{\alpha}^{*} d_{\alpha}^{ }}
\def\na{n_{\alpha}^{ }}
\def\nua{\nu_{\alpha}^{ }}
\def\nuab{\nu_{{-\alpha}}^{ }}
\def\qa{q_{\alpha \sigma}^{ }}
\def\qab{q_{{-\alpha}\sigma}^{ }}
\def\qaqa{q_{\alpha \sigma}^{2}}
\def\qaqab{q_{{-\alpha}\sigma}^{2}}

\def\mbas{\bar{m}_{\alpha'}^{ }}
\def\xibas{\bar{\xi}_{\alpha'}^{ }}
\def\xibabs{\bar{\xi}_{{-\alpha'}}^{ }}
\def\das{d_{\alpha'}^{ }}
\def\cdas{d_{\alpha'}^{*}}
\def\cdadas{d_{\alpha'}^{*} d_{\alpha'}^{ }}
\def\nas{n_{\alpha'}^{ }}
\def\nuas{\nu_{\alpha'}^{ }}
\def\nuabs{\nu_{{-\alpha'}}^{ }}
\def\qas{q_{\alpha'\sigma}^{ }}
\def\qabs{q_{{-\alpha'}\sigma}^{ }}
\def\qaqas{q_{\alpha'\sigma}^{2}}
\def\qaqabs{q_{{-\alpha'}\sigma}^{2}}

\def\mbass{\bar{m}_{\alpha''}^{ }}
\def\xibass{\bar{\xi}_{\alpha''}^{ }}
\def\xibabss{\bar{\xi}_{{-\alpha''}}^{ }}
\def\dass{d_{\alpha''}^{ }}
\def\cdass{d_{\alpha''}^{*}}
\def\cdadass{d_{\alpha''}^{*} d_{\alpha''}^{ }}
\def\nass{n_{\alpha''}^{ }}
\def\nuass{\nu_{\alpha''}^{ }}
\def\nuabss{\nu_{{-\alpha''}}^{ }}
\def\qass{q_{\alpha''\sigma}^{ }}
\def\qabss{q_{{-\alpha''}\sigma}^{ }}
\def\qaqass{q_{\alpha''\sigma}^{2}}
\def\qaqabss{q_{{-\alpha''}\sigma}^{2}}

\def\qb{q_\sigma^{o}}
\def\mnu{\nu_{ }^{o}}
\def\mxi{\xi_{ }^{o}}

\def\qba{\langle q_{\alpha \sigma}^{ }\rangle}
\def\mnua{\langle \nu_{\alpha}^{ }\rangle}
\def\mxia{\langle \alpha \bar{\xi}_{\alpha}^{ }\rangle}

\def\IKso{\langle s_0^{ } \rangle}
\def\IKsl{\langle s_l^{ } \rangle}
\def\IKsosl{\langle s_0^{ }s_l^{ } \rangle}
\def\bPsi{\bar{\mit \Psi}}
\def\bcS{\bar{\cS}}

\def\Kh{K_{\bar{h}}^{ }}
\def\KJ{K_{\bar{J}}^{ }}
\def\Sh{S_{\bar{h}}^{ }}
\def\SJ{S_{\bar{J}}^{ }}
\def\Tha{\mbox{Th}_{\alpha}^{ }}
\def\tanha{\tanh{[\beta(\mbox{$\frac{1}{z}$}\bar{h}+h^*+\alpha \bar{J})]}}

\def\GOii{G_{ii\sigma}^{o}}
\def\GOij{G_{ij\sigma}^{o\prime}}
\def\GOji{G_{ji\sigma}^{o\prime}}

\input{epsf}
\begin{abstract}
The uniform static spin susceptibility in the paraphase of the one--band 
Hubbard model is calculated within a theory of magnetic short--range order 
(SRO) which extends the four--field slave--boson functional--integral approach
by the transformation to an effective Ising model and the self--consistent 
incorporation of SRO at the saddle point. This theory describes a 
transition from the paraphase without SRO for hole dopings 
$\delta > \delta_{c_2}$ to a paraphase with antiferromagnetic SRO for 
$\delta_{c_1} < \delta < \delta_{c_2}$. In this region 
the susceptibility consists of interrelated 
`itinerant' and `local' parts and increases upon doping. 
The zero--temperature susceptibility exhibits a cusp at $\delta_{c_2}$
and  reduces to the usual slave--boson result for larger dopings. 
Using the realistic value of the on--site Coulomb repulsion $U=8t$ for 
$\rm La_{2-\delta}Sr_{\delta}CuO_{4}$, 
the peak position ($\delta_{c_2} = 0.26$) as well as the doping 
dependence reasonably agree with low--temperature susceptibility experiments 
showing a maximum at a hole doping of about 25\%.
\end{abstract}
\parindent0.8cm
{\bf PACS number(s):} {\it 75.10.-b, 71.28.+d, 71.45.-d} 
\thispagestyle{empty}

\vspace*{1.5cm}
Among the most striking features of high--$T_c$
superconductors in the normal state, the 
unconventional magnetic properties have attracted
increasing attention \cite{KA94}.
As revealed by neutron scattering \cite{RMea93a} and
nuclear magnetic resonance~\cite{Hoea90} experiments, in the metallic state
there exist pronounced antiferromagnetic (AFM) spin correlations which are
ascribed to strong Coulomb correlations within the $\rm CuO_2$ planes. 
Knight--shift~\cite{SKPH93} and bulk
measurements~\cite{SKPH93,Toea89,Jo89}
of the spin susceptibility ${\mit \chi}(T,\delta)$ in
$\rm La_{2-\delta}Sr_\delta CuO_4$ show a maximum in the doping
dependence as well as, for moderate hole doping ($\delta \le 0.21$), 
in the temperature dependence, 
where the temperature of the maximum decreases with increasing
doping. Such a behaviour, also observed in $\rm YBa_2Cu_3O_{6+y}$ ($y\le
0.92$)~\cite{HD93,Waea90},
 may be qualitatively understood as an 
effect of AFM short--range order (SRO) which decreases with increasing
doping and temperature.

Up to now there have been only a few attempts, based on
one--band~\cite{Tr90a,TKF91,KM91,MMM95} 
and three--band \cite{BSB93} 
correlation models, to describe the unusual doping and temperature 
dependence of the normal--state susceptibility. In the t--t'--J model, a
maximum in $\mit \chi$ was obtained for the Pauli susceptibility of a
strongly renormalized quasiparticle band \cite{Tr90a} or for the
RPA slave--boson susceptibility~\cite{TKF91} showing a cusp in the temperature
dependence at the transition to the singlet RVB state. In the one-band Hubbard
model, a maximum in the doping dependence of $\chi$ was found by a
semi-phenomenological weak--coupling approach \cite{KM91} or by the composite
operator method \cite{MMM95}. The role played by SRO in explaining 
the normal--state susceptibility was investigated on the basis of 
the three--band Hubbard model~\cite{BSB93} by means of a slave--boson CPA
 theory which, however, is self--consistent only at the single--site
level and does not hold at very low temperatures. To improve the treatment of
SRO in the paraphase being valid also at $T=0$, in a previous 
communication \cite{TIF95a}, hereafter referred to as I, we have
presented the main
features of a theory of magnetic SRO in the one--band Hubbard model based on 
the scalar four--field  slave--boson (SB) approach \cite{KR86}. 
In I we have focused on 
the stability of magnetic long--range order (LRO) versus SRO, where magnetic
LRO phases are found to make way to a paraphase with SRO in a wide doping 
region.

In this Brief Report we extend our theory by the inclusion of an
external magnetic 
field $h$  and by the calculation of the uniform static spin
susceptibility $\chi$ in
the paraphase, where special care is taken to the influence of SRO.

Following the lines indicated in I, the action of the SB functional integral
for the partition function of the 2D Hubbard model is expressed in terms of the
SB fields $m_{i}$, $\xi_{i}$, $n_{i}$, $\nu_{i}$, $d_{i}$ and
$d_{i}^{*}$ \cite{Ha89}. To
treat the fluctuations of the local magnetizations $m_{i}$ and the internal
magnetic fields $\xi_{i}$ we write $m_i^{ } = \mb_i^{ }\si$, 
$\xi_i^{ } = \xib_i^{ }\si$ ($\si = \pm$) and make the ansatz
$b_i \to b_{s_i}$ for the magnetic amplitudes $b \in \{\mb, \xib \}$ and
the charge degrees of freedom $b \in \{ n , \nu, d=d_{ }^{*} \}$. We
transform the free--energy functional $\mit \Psi$ to an effective Ising model
in the nearest--neighbour pair ($\langle ij \rangle$) approximation and obtain
\begin{equation}
\label{PsiIs}
{\mit \Psi}(\{s_i\}) = \bPsi -\bar{h} \sum_i \si 
- \bar{J} \sum_{\langle ij \rangle} \si \sj\;,
\end{equation}
with
\begin{eqnarray}
\label{barPSI}
\bPsi &=& \mbox{$-\frac{1}{\beta}$}
\sum_{\vec{k}\sigma}
\ln{\left[1+\exp{\{-\beta[(z_{\sigma}^{o})_{ }^2 \varepsilon_{\vec{k}}
+ \mnu -\sigma(\mxi+h)-\mu]\}}
\right]} \nonumber \\
& & +\mbox{$\frac{N}{2}$}
\sum_{\alpha=\pm 1}
\left\{ U d_{\alpha}^{2} -\na \nua +\mba \xiba + \sum_{\sigma}
\left({\mit \Phi}_{\alpha \sigma}^{ }
     +{\mit \Phi}_{\alpha \alpha \sigma}
     +{\mit \Phi}_{{-\alpha} \alpha \sigma}\right)\right\}\;, \\
\bar{h} &=& -\mbox{$\frac{1}{2}$} \sum_{\alpha}\alpha
\left[U d_{\alpha}^{2} -\na \nua +\mba \xiba
 + \sum_{\sigma} \left( {\mit \Phi}_{\alpha \sigma}
                    +2 {\mit \Phi}_{\alpha \alpha \sigma}
                    \right)\right]\;, \\
\label{JIsing}
\bar{J} &=& -\mbox{$\frac{1}{4}$}\sum_{\alpha \sigma}\left(
                 {\mit \Phi}_{\alpha \alpha \sigma}
                -{\mit \Phi}_{{-\alpha} \alpha \sigma} \right) \;.
\end{eqnarray}
The single--site and two--site fluctuation
contributions 
${\mit \Phi}_{\alpha \sigma}^{ }=\left.{\mit \Phi}_{i\sigma}^{
}(\alpi)\right|_{\alpi=\alpha}$ and
${\mit \Phi}_{\alpha \alpha^{\prime}\sigma}=\left.
{\mit \Phi}_{\langle ij\rangle \sigma}(\alpi,\alpj)
\right|_{{\alpi=\alpha}^{ }\atop {\alpj=\alpha^{\prime}}}$,
respectively, are given by
\begin{eqnarray}
\label{phia}
{\mit \Phi}_{i\sigma}^{ } &=& 
\mbox{$\frac{1}{\pi}$} \int d\omega f(\omega - \mu)
\mbox{Im}~\ln{\left[1-\GOii V_{i\sigma}^{ }(\alpha_i^{ })\right]} \;,\\
\label{phiaa}
{\mit \Phi}_{\langle ij \rangle \sigma}^{ } &=& \mbox{$\frac{1}{\pi}$}
\int d\omega f(\omega - \mu)
\mbox{Im}~\ln\left[1-G_{\langle ij\rangle\sigma}^{o}
T_{j\sigma}(\alpj)G_{\langle ji\rangle\sigma}^{o}
T_{i\sigma}(\alpi)\right]\;.
\end{eqnarray}
In (\ref{phiaa}), $G_{ij\sigma}^{o}(\omega)$ is the uniform
paramagnetic (PM) Green propagator, and 
the scattering matrix $T_{i\sigma}=V_{i\sigma}^{ }\,
\left(1-\GOii V_{i\sigma}^{ }\right)^{-1}$ is expressed in terms of the
local perturbation 
\begin{equation}
V_{i\sigma}^{ }(\alpha_i,\omega) 
=\frac{1}{(z_{\alpha_{i}^{ }\sigma}^{ })^2}\left\{
\left[(z_{\alpha_{i}^{ }\sigma}^{ })^2-(z_{\sigma}^{o})^2\right]
\left[\omega - \mnu +\sigma(\mxi+h)\right]
+(z_{\sigma}^{o})^2\left[\nu_{\alpha_i}^{ }-\mnu - 
\sigma (\alpha_i \bar{\xi}_{\alpha_i}^{ }-\mxi)\right] \right\}\;,
\end{equation}
where
\begin{eqnarray}
z_{\alpha_{i}^{ }\sigma}^{ }
&=&\frac{\sqrt{2}\left[\sqrt{(n_{\alpha_{i}^{ }}^{ }
+\sigma \alpha_i \bar{m}_{\alpha_{i}^{ }}^{ }-2d_{\alpha_{i}^{ }}^2)
(1-n_{\alpha_{i}^{ }}^{ }+d_{\alpha_{i}^{ }}^2)}+d_{\alpha_{i}^{ }}^{ }
\sqrt{n_{\alpha_{i}^{ }}^{ }-\sigma \alpha_i \bar{m}_{\alpha_{i}^{ }}^{ } 
-2d_{\alpha_{i}^{ }}^2}\,\right]}
{\sqrt{(n_{\alpha_{i}^{ }}^{ }+\sigma \alpha_i 
\bar{m}_{\alpha_{i}^{ }}^{ })
(2 - n_{\alpha_{i}^{ }}^{ }-\sigma \alpha_i \bar{m}_{\alpha_{i}^{ }}^{ })}}\;,
\end{eqnarray}
and the superscript ``$o$''
 refers to PM saddle--point values. By the functional (\ref{PsiIs})
we determine the saddle point for all Bose fields 
$b_{\alpha}= \{\mb_{\alpha}, \xib_{\alpha} ,n_{\alpha} , \nu_{\alpha},
d_{\alpha}\}$ in the external field $h$, where, in the spirit of I, the SRO is
self--consistently incorporated within the Bethe cluster approximation (taking
into account only the nearest--neighbour SRO). As found in I, in the $h=0$
limit ($\mba=\mb$) the self--consistent calculation of the effective
Ising--exchange integral $\bar{J}$ as function of the interaction strength $U$
and the hole doping $\delta=1-n$ yields two possible paraphases ($\langle
s_{i} \rangle=0$): (i) the paraphase without SRO (PM; $\bar{J}=0$, $\mb=0$) and
(ii) the paraphase with antiferromagnetic SRO (SRO--PM; $\bar{J}<0$, $\mb>0$).

The uniform static spin susceptibility $\mit \chi(T,\delta)$ has to be
calculated according to
\begin{equation}
\label{chidef}
{\mit \chi}=\lim_{h\to 0}\,\sum_\alpha \left( W_\alpha \frac{d\,m_\alpha}{d\,h}
+m_\alpha \frac{d\,W_\alpha}{d\,h}\right)\;,
\end{equation}
where $m_{\alpha}=\mba \alpha$, $W_{\alpha}=W_{\alpha}(\bar{h},
h_{ }^{*},\bar{J})$
is the probability for the Ising spin $\alpha$ at the central site of the
Bethe cluster, and $h^{*}$ is the effective Bethe field. The first term in
(\ref{chidef})  describes the change of the
magnetization--amplitude with the applied magnetic field and gives
mainly the `itinerant' contribution to $\chi$. The second term
describes directional fluctuations of the local magnetizations and is
called the `local' contribution being finite only in the SRO--PM phase.
Note that the `itinerant' and `local' properties are interrelated
and determine {\it both} contributions to the spin susceptibility.
In the PM and SRO-PM phases we have calculated the doping dependence of the zero--temperature
susceptibility in the 2D Hubbard model (being finite in contrast to the theory
of Ref.\cite{BSB93}) in a completely self--consistent way, where in 
the tedious numerical evaluation of 
the integrals~(\ref{phia}) and~(\ref{phiaa}) and of their 
derivatives 
particular attention has to be paid to the analytical
behaviour of the complex logarithm.

\begin{figure}
 \centerline{\mbox{\epsfxsize 13cm\epsffile{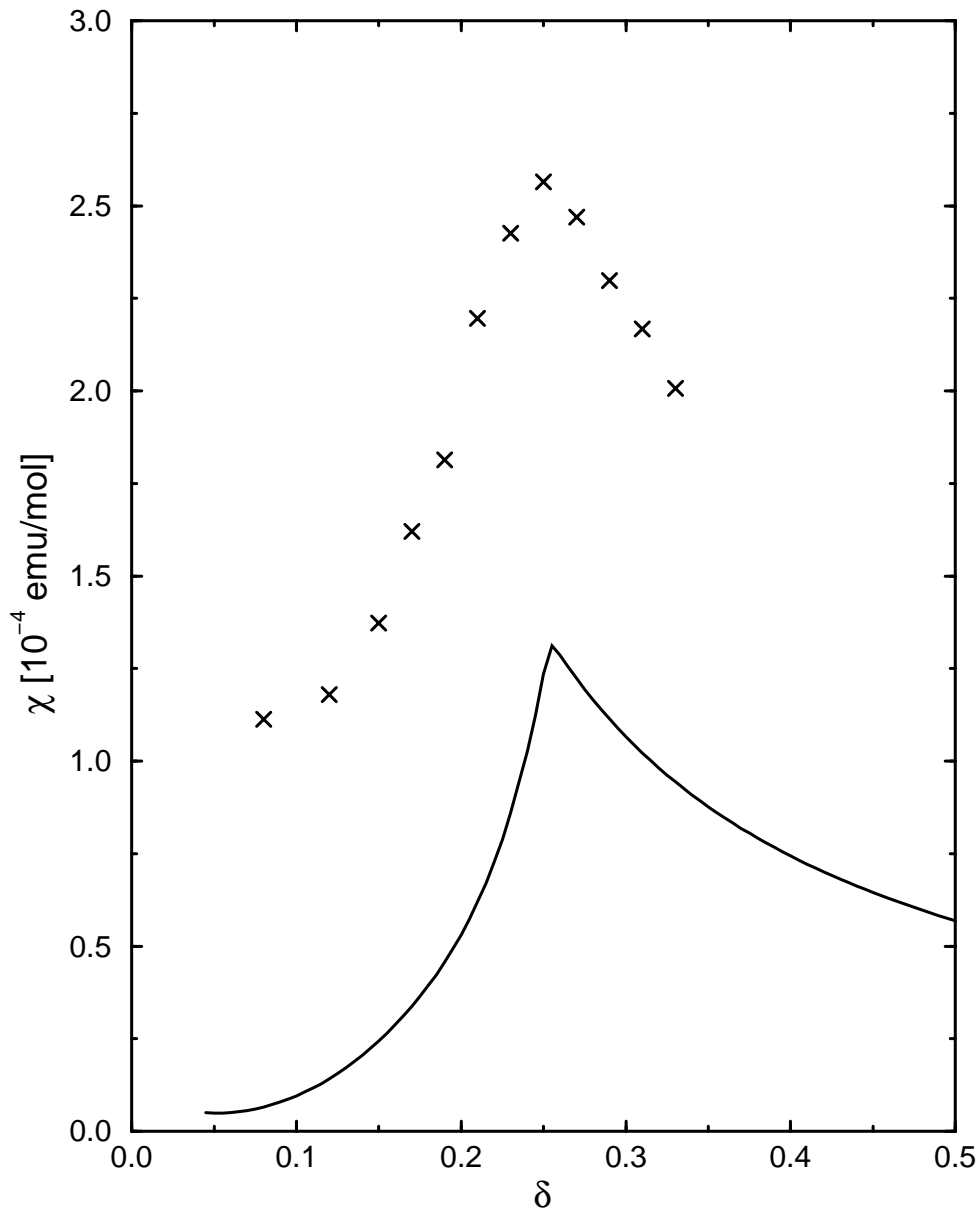}}}
 \caption{Uniform static spin susceptibility as a function of
  doping at $T=0$. The theoretical result obtained for the  2D Hubbard 
  model at $U/t=8$ and  $t=0.3$~eV (solid) is compared with the spin
  contribution ($\times$) to the (corrected) experimental susceptibility 
   on $\rm La_{2-\delta}Sr_{\delta}CuO_4$
  at $T=50$~K~\protect\cite{Toea89,Jo89}.} 
 \end{figure}

Figure~1 shows our result without any fit procedure 
using the commonly
accepted value $U/t=8$ for the Hubbard model applied to high--$T_c$
cuprates~\cite{HSSJ90}. As stated in I,
in the region 
$6 < U/t<12$, there occurs a
first--order (1,1)--spiral$\rightleftharpoons$SRO--PM transition at
$\delta_{c_1}$ and a SRO--PM$\rightleftharpoons$PM 
transition of second order at $\delta_{c_2}$.
In the PM phase ($\delta>\delta_{c_2}$) the SB band--renormalized Pauli
susceptibility has a pronounced doping dependence in two
dimensions and agrees with the static and uniform limit of the dynamic spin
susceptibility derived, within the spin--rotation--invariant 
SB scheme~\cite{LWH89}, from
the Gaussian fluctuation matrix at the PM saddle point~\cite{LSW91}.
In the SRO--PM phase ($\delta_{c_1}<\delta<\delta_{c_2}$),
the Pauli susceptibility is suppressed due to
the SRO--induced spin stiffness against the orientation of the local
magnetizations along the
homogeneous external field. Accordingly, at $\delta_{c_2}$ a cusp in
$\chi(0,\delta)$ appears. Since, for
$\delta_{c_1}<\delta<\delta_{c_2}$, 
$|\bar{J}|$ decreases with increasing $\delta$~\cite{TIF95a}, 
the susceptibility increases upon doping.

The peak in $\chi(0,\delta)$ only appears at sufficiently high ratios
$U/t > 6$, for which a SRO--PM$\rightleftharpoons$PM transition may occur.
According to the phase diagram, given in
Fig.~2 of I, in the region $6<U/t<12$ the
SRO--PM$\rightleftharpoons$PM transition shifts to higher doping
values with increasing $U/t$. Correspondingly, the peak position in
$\chi(0,\delta)$ reveals the same $U/t$ dependence.

In Fig.~1 we have also depicted the spin contribution to the magnetic 
susceptibility of 
$\rm La_{2-\delta}Sr_\delta CuO_4$ at 50~K
obtained from the experimental data on the total susceptibility \cite{Toea89}
by subtracting the diamagnetic core ($-9.9 \times
10^{-5}$ emu/mol) and Van Vleck ($ 2.4 \times 10^{-5}$ emu/mol) 
contributions which, according to Ref.~\cite{Jo89}, can be taken to be 
independent of doping and temperature 
over the limited parameter region studied here. As Fig.~1 shows,
the experimentally observed pronounced maximum at a hole doping of about 25\%
is reproduced very well by our theory yielding the peak position at
$\delta_{c_2}=0.26$ ($U/t=8$). Moreover, the qualitative doping dependence
of $\chi$ reasonably agrees with experiments.
Of course, it could not be expected that
our approach based on the simple (single--band) Hubbard model yields the
correct magnitude of $\chi$ for $\rm La_{2-\delta}Sr_\delta CuO_4$.
Especially, concerning the low--doping limit $\delta \to
\delta_{c_1}^{ }=0.04$, the theoretical susceptibility
is much too low as compared with experiments. This deficiency may be explained
as follows. For $\delta=0$ and large $U/t$ values, the Hubbard model is
equivalent to the Heisenberg antiferromagnet with the exchange interaction
$J=4t_{ }^{2}/U$. In this model, the spin susceptibility at $T=0$ has a finite
value proportional to $J_{ }^{-1}$ \cite{Ba91} which is due to the existence
of transverse spin fluctuations. However, our {\it scalar} four--field SB
approach to the spin susceptibility in the presence of SRO implies the
transformation of the free--energy functional to an effective Ising model
describing longitudinal fluctuations only. 
Since the `local' contribution to $\chi$ is of Ising-type, we get a
too small susceptibility in the low--doping limit which, however, is
finite due to the interrelation to the `itinerant' contribution to $\chi$.
Therefore, we suggest that a
theory of SRO based on the spin--rotation--invariant SB scheme \cite{LWH89}
and resulting in an effective Heisenberg--model functional may improve the
results in the magnitude of $\chi$, in particular at low doping levels.

Finally, we notice that the increase of the susceptibility upon doping
obtained within our theory for moderate Coulomb repulsions 
($U/t>6$) is in qualitative accord with recent QMC data~\cite{CT94} 
and with the approaches of Refs. \cite{KM91} and \cite{MMM95}.
However, in those works a maximum in the spin susceptibility 
was found even at a smaller coupling ($U/t=4$).

From our results we conclude that the concept of magnetic SRO in
strong--correlation models may play the key role in the explanation of
many unconventional properties of high--$T_c$ compounds.
The theory may be extended in several directions. As discussed above a
spin--rotation--invariant theory of SRO may improve the agreement of the spin
susceptibility with experiments. Furthermore, as
motivated by neutron scattering experiments~\cite{RMea93a} 
probing the AFM correlation length over several lattice spacings, the 
effects of a longer than nearest--neighbour ranged SRO (which may be described
beyond the nearest--neighbour pair approximation) should be
investigated.

This work was performed under the auspices of
Deutsche For\-schungsgemeinschaft under project SF--HTSL--SRO.
H.F. acknowledges the support of the NTZ and 
the hospitality at the University of Leipzig.
\baselineskip0.62cm
\bibliography{ref}

\begin{thebibliography}{10}

\bibitem{KA94}
A.~P. Kampf, Physics Reports {\bf 249}, 219 (1994).

\bibitem{RMea93a}
J.~Rossat-Mignod et~al., Physica B {\bf 186--188}, 1 (1993).

\bibitem{Hoea90}
M.~Horvatic et~al., Physica C {\bf 166}, 151 (1990).

\bibitem{SKPH93}
Y.-Q. Song, M.~A. Kennard, K.~R. Poeppelmeier, and W.~P. Halperin, Phys. Rev.
  Lett. {\bf 70}, 3131 (1993).

\bibitem{Toea89}
J.~B. Torrance et~al., Phys. Rev. B {\bf 40}, 8872 (1989).

\bibitem{Jo89}
D.~C. Johnston, Phys. Rev. Lett. {\bf 62}, 957 (1989).

\bibitem{HD93}
Z.~P. Han and R.~Dupree, Physica C {\bf 208}, 328 (1993).

\bibitem{Waea90}
R.~E. Walstedt et~al., Phys. Rev. B {\bf 41}, 9574 (1990); 
B~{\bf 45}, 8047 (1992).

\bibitem{Tr90a}
S.~A. Trugman, Phys. Rev. Lett. {\bf 65}, 500 (1990).

\bibitem{TKF91}
T.~Tanamoto, K.~Kuboki, and H.~Fukuyama, J. Phys. Soc. Jpn. {\bf 60}, 3072
  (1991); T.~Tanamoto, H.~Kohno, and H.~Fukuyama, 
J. Phys. Soc. Jpn. {\bf 62}, 717
  (1993); {\bf 62}, 1455 (1993).

\bibitem{KM91}
T.~Kopp and F.~Mila, Phys. Rev. B {\bf 43}, 12980 (1991); 
{\bf 50}, 13017 (1994).

\bibitem{MMM95}
F.~Mancini, S.~Marra, and H.~Matsumoto, Physica C {\bf 252}, 361 (1995).

\bibitem{BSB93}
G.~Baumg\"artel, J.~Schmalian, and K.~H. Bennemann, Europhys. Lett. {\bf 24},
  601 (1993).

\bibitem{TIF95a}
U.~Trapper, D.~Ihle, and H.~Fehske, Phys. Rev. B {\bf 52}, R11553 (1995).

\bibitem{KR86}
G.~Kotliar and A.~E. Ruckenstein, Phys. Rev. Lett. {\bf 57}, 1362 (1986).

\bibitem{Ha89}
H.~Hasegawa, J. Phys. Condens. Matter {\bf 1}, 9325 (1989).

\bibitem{HSSJ90}
M.~S. Hybertsen, E.~B. Stechel, M.~Schl\"uter, and D.~R. Jennison, Phys. Rev. B
  {\bf 41}, 11068 (1990); M.~S. Hybertsen, 
E.~B. Stechel, W.~M.~C. Foulkes, and M.~Schl\"uter, Phys. Rev.
  B {\bf 45}, 10032 (1992).

\bibitem{LWH89}
T.~Li, P.~W\"olfle, and P.~J. Hirschfeld, Phys. Rev. B {\bf 40}, 6817 (1989).

\bibitem{LSW91}
T.~Li, Y.~S. Sun, and P.~W\"olfle, Z. Phys. B {\bf 82}, 369 (1991).

\bibitem{Ba91}
T.~Barnes, Int. J. Mod. Phys. C {\bf 2}, 659 (1991).

\bibitem{CT94}
L.~Chen and A.-M.~S. Tremblay, Phys. Rev. B {\bf 49}, 4338 (1994).

\end{thebibliography}
\bibliographystyle{phys}
 \end{document}